\documentclass[aip,graphicx,amsmath,amssymb,reprint]{revtex4-1}
\usepackage{times}
\usepackage{natbib}
\usepackage{stix}
\usepackage{amsmath,amssymb}

\usepackage{graphicx}
\usepackage{color}

\newcommand\Alfven{Alfv\'en }

\newcommand{\T}[1]{\texttt{#1}} 
\newcommand{\V}[1]{\mathbf{#1}}

\newcommand{\bhat}{\mbox{$\hat{\mathbf{b}}$}}
\newcommand{\figref}[1]{Fig.~\ref{#1}}   
   
\begin{document}


\title[Plasma Seismology]{Plasma Seismology: Fully Exploiting the Information Contained in Velocity Space of Kinetic Plasmas using the Morrison $G$ Transform}
\author{F.~Skiff}
 \email{frederick-skiff@uiowa.edu}
\author{G.~G.~Howes}%
\affiliation{ Department of Physics and Astronomy, University of Iowa, Iowa City, IA 52245}

\date{\today}

\begin{abstract}
Weakly collisional plasmas contain a wealth of information about the dynamics of the plasma in the particle velocity distribution functions, yet our ability to exploit fully that information remains relatively primitive.  Here we aim to present the fundamentals of a new technique denoted Plasma Seismology that aims to invert the information from measurements of the particle velocity distribution functions at a single point in space over time to enable the determination of the electric field variation over an extended spatial region.  The fundamental mathematical tool at the heart of this technique is the Morrison $G$ Transform.  Using kinetic numerical simulations of Langmuir waves in a Vlasov-Poisson plasma, we demonstrate the application of the standard Morrison $G$ Transform, which uses measurements of the particle velocity distribution function over all space at one time to predict the evolution of the electric field in time.  Next, we introduce a modified Morrison $G$ Transform which uses measurements of the particle velocity distribution function at one point in space over time to determine the spatial variation of the electric field over an extended spatial region.  We discuss the limitations of this approach, particularly for the numerically challenging case of Langmuir waves. The application of this technique to \Alfven waves in a magnetized plasma holds the promise to apply the technique to existing spacecraft particle measurement instrumentation to determine the electric fields over an extended spatial region away from the spacecraft.
\end{abstract}


\maketitle 

\section{Introduction}

Helioseismology has proven to be a groundbreaking technological innovation, 
making possible a determination of the internal structure and dynamics of the Sun
using only observations of the vibrational modes at the surface of the Sun. 
Here we aim to develop an analogous innovation, denoted \emph{Plasma Seismology}, 
using measurements of the velocity distribution function (VDF) of particles at a single point in 
space to reconstruct the variation of the electric field over an extended spatial region.
A successful implementation of this technique on spacecraft will enable far greater information 
about the electric field variation in space to be deduced from VDF measurements made with 
existing particle instrumentation (\emph{e.g.}, electrostatic analyzers).

Under the weakly collisional conditions typical of space and astrophysical plasmas, as well as of many laboratory plasmas, collisionless interactions between the electromagnetic fields and plasma particles dominate the energy transport, plasma heating, and particle acceleration that govern the long-term evolution of the system.  Such fundamental interactions underlie some long-standing questions in heliophysics, such as determining how the solar corona is heated to more than a million Kelvin (nearly 1000 times hotter than the solar photosphere), or how solar flares accelerate ions and electrons to nearly the speed of light that cause solar-energetic-particle (SEP) events that can harm our technological infrastructure at Earth, including navigation and communication satellites.

The key to unraveling the physical mechanisms governing these collisionless field-particle interactions lies in fully exploiting the information contained in the six-dimensional (3D-3V) phase space of kinetic plasma physics. In particular, understanding how the particle velocity distribution functions (VDFs) evolve in these interactions provides the ability both to identify the mechanisms at play and to determine the rate of energization of the different populations of particles in velocity space, providing a concrete path to answering a wide range of important science questions \citep{Howes:2018b}.  Analogous to how helioseismology combines knowledge of the properties of sound waves in plasma with  the observed vibrational modes at the surface of the Sun to infer its internal structure and dynamics, \emph{plasma seismology} combines knowledge of how particles respond to electromagnetic fields  with the observed particle velocity distributions at one position to infer the variation of the electric field over an extended spatial region.

Successfully interpreting the information contained in the particle velocity distributions is challenging due to the high-dimensionality of 
the six-dimensional (3D-3V) phase space of kinetic plasmas, the large computational cost for 3D-3V simulations, and the difficulty of experimentally or observationally measuring the velocity distributions, particularly in laboratory plasmas.  However, recent advancements in all of these areas have lead to several significant successes in recent years.  For example, ion kinetic instabilities play a key role in the thermodynamic evolution of the solar wind plasma \citep{Bale:2009,Alexandrova:2013b}, and recent work has successfully applied Nyquist's criterion for kinetic stability \citep{Nyquist:1932} to ion velocity distribution functions (iVDFs)  measured by spacecraft \citep{Klein:2017c}
to show that nearly half of a sample of  309 intervals are linearly unstable \citep{Klein:2018}; further work using machine learning to analyze over one million intervals enables the instability dynamics to be identified based on the observed iVDF \citep{Klein:2019,Martinovic:2021,Martinovic:2023}, demonstrating clearly that particle VDFs conceal a vast store of information on the system dynamics. 

Another example of exploiting velocity space is the recently developed field-particle correlation (FPC) technique \citep{Klein:2016a,Howes:2017a,Klein:2017b}---an approach that combines both electromagnetic field and particle velocity distribution measurements---which has been used with \emph{Magnetospheric Multiscale} (\emph{MMS}) spacecraft measurements to demonstrate that electron Landau damping plays an important role in the dissipation of turbulence in Earth's magnetosheath plasma \citep{Chen:2019,Afshari:2021}. The FPC technique was also applied to experiments on the Large Plasma Device (LAPD) \citep{Gekelman:1991,Gekelman:1999} that demonstrated successfully in the laboratory that electrons can be accelerated by \Alfven waves under conditions of the Earth's auroral magnetosphere \citep{Schroeder:2021},  experimentally confirming a more than 40-year-old hypothesis about one source of precipitating auroral electrons \citep{Hasegawa:1976c}.

These examples make clear that the development of new techniques to make full use of the large amount of information contained in the particle VDFs holds the promise to open new avenues for making substantial progress in unraveling the kinetic physics of weakly collisional plasmas.  
The concept of plasma seismology---taking advantage of the nonlocal nature of particle trajectories in phase space to employ the measured particle VDFs at one position to reconstruct the electric field over an extended spatial extent---is based on the mathematics of the \emph{Morrison $G$ transform} \citep{Morrison:1992,Morrison:1994,Morrison:2000,Heninger:2018}. This technique has been explored in preliminary experiments \citep{Sarfaty:1996,Skiff:1987,Skiff:1992,Skiff:2000,Skiff:2002a,Skiff:2002b}, but with the experimental diagnostics and numerical tools we have at hand, the time is right to extend our work to exploit the Morrison $G$ transform to develop the innovative technique of plasma seismology.  

In Sec.~\ref{sec:morrison}, we review the mathematical formulation of the standard Morrison $G$ Transform and introduce the 
Morrison $\widetilde{G}$ Transform that serves as the foundation of the plasma seismology technique.  We describe kinetic simulations of Langmuir waves using the Nonlinear Vlasov-Poisson Code \T{VP} \citep{Howes:2017a} that we employ to test the plasma seismology reconstruction of the spatial variation of the electric field in Sec.~\ref{sec:sims}.  The results of applying both the 
standard and modified versions of the transform to the simulation data are presented in Sec.~\ref{sec:results}, and the uncertainties and limitations of the plasma seismology technique are detailed in Sec.~\ref{sec:limitations}.  In Sec.~\ref{sec:conc}, we conclude by highlighting the future direction of this research.

\section{The Morrison $G$ Transform}
\label{sec:morrison}
Here we review the mathematical formulation of the standard Morrison $G$ Transform \citep{Morrison:1992,Morrison:1994,Morrison:2000,Heninger:2018}, which is cast as an initial value problem.
Then we introduce a complementary formulation of the Morrison $G$ Transform cast instead as a boundary value problem, denoted here as the modified Morrison $\widetilde{G}$ Transform, with the tilde to distinguish these two forms.  This latter formulation represents the mathematical framework underpinning the novel technique of \emph{plasma seismology} introduced here.

\subsection{The Standard Morrison $G$ Transform}
We use the term "standard Morrison $G$ Transform" to refer to the diagonalizing transform introduced by Morrison and colleagues to project the linearized Vlassov-Poisson equation onto the basis of its eigenfunctions---the Case-Van Kampen modes
\citep{Morrison:1992,Morrison:1994,Morrison:2000,Heninger:2018}.  Here we will derive the transform by the introduction of three functions of a single complex variable ($v$) related by multiplication,
\begin{equation}
\tag{1}
  F=G*\epsilon  
\end{equation}  
all of which are analytic in the upper half plane of $v$.  By Hilbert's theorem, this means that the the real and imaginary parts of each of these functions evaluated on the real axis are related by a Hilbert transform.  One must subtract off the values of the functions at the point $v=\infty$.  For example one has 
\begin{equation}
    Real(\epsilon(v)-\epsilon(\infty))= Hilbert[Imag(\epsilon(u)-\epsilon(\infty))]_v
\end{equation}
with $v$ restricted to the real axis.  We follow Morrison in defining the Hilbert transform as
\begin{equation}
 Hilbert[f(u)]_v \equiv\frac{p}{\pi}\int\frac{f(u)}{(u-v)}du. 
\end{equation}
The letter p indicates that the singular integral is to be evaluated in the sense of the Cauchy Principal value.  Applying the Hilbert transform twice returns the original function with a multiplicative factor of (-1).  If we consider $f$ and $g$ to be the imaginary parts of $F$ and $G$ respectively along the real axis, then the following relations between $f$ and $g$ follow immediately from Equation (1)
\begin{equation}
g(u)=\frac{Real(\epsilon)f(u)}{|\epsilon|^2}-\frac{Imag(\epsilon)Hilbert[f(v)]_u}{|\epsilon|^2}
\end{equation}
which we can write as $g=\hat{G}[f]$ and
\begin{equation}
f(v)=Real(\epsilon)g(v)+Imag(\epsilon)Hilbert[g(u)]_v
\end{equation}
 (or $f=G[g]$). If the function $ \epsilon$ is taken to be the plasma dielectric function, then we say that $f$ is the Morrison transform of $g$.  In the classic case of the electron plasma wave we have
\begin{equation}
Imag(\epsilon)=-\pi\frac{\omega_{p}^2}{k^2}\frac{\partial f_0}{\partial v}
\end{equation}
where $\omega_p^2=\frac{n_0 e^2}{\varepsilon_0 m_e}$ introduces the electron plasma frequency, $k$ is the wave number, and $f_0$ is the equilibrium zeroth-order electron velocity distribution function normalized such that
\begin{equation}
\int f_0(v) dv = 1.
\end{equation}
The real part of $\epsilon$ is equal to its value at $v=\infty$ (which is 1) added to the Hilbert transform of the imaginary part.  When interpreted as a dispersion relation $\epsilon (\omega,k)=0$ the complex variable $v$ is replaced by $\omega/k$ and the complex $v$-plane is equivalent to the complex $\omega -$plane.  However, for the G-transformation, only values of $\epsilon (v)$ with a real argument are required. What is both remarkable and useful is that the inverse Morrison transform applied to the linearized Vlasov-Poisson equation for electron plasma waves ($f$ being the dimensionless small perturbation to $f_0$)
\begin{equation}
    \frac{\partial f}{\partial t}+ikvf+\frac{\omega_p^2}{ik}\frac{\partial f_0}{\partial v}\int f dv =0
\end{equation}
produces a simple equation in $g$ that shows that $g(u)$ is the amplitude of the CVK component with phase velocity $u$.
\begin{equation}
    \frac{\partial g}{\partial t} +ikug=0
\end{equation}
It is well known that a general solution of the linear initial-value problem for the Vlasov-Poisson equation can be given in terms of CVK modes.  In using the G-transform approach we see that there exists a family of CVK spectra - one for each value of k 
\begin{equation}
    g(u,x,t)=\sum_k g_0(u,t=0)_k e^{ikx-ikut}.
\end{equation}
If we express the initial conditions on $f$ using a discrete sum over $k$ (appropriate for the numerical solutions we will use for illustration)
\begin{equation}
    f(x,v,t=0)=\sum_k f(v,t=0)_k e^{ikx}
\end{equation}
Then we can determine the initial $g's$ from the initial $f's$.
\begin{equation}
g_0(u,t=0)_k = \hat{G}_k[f(v,t=0)_k]
\end{equation}
Because our numerical simulation keeps nonlinear effects, one expects that the CVK analysis will have a range of validity at low wave amplitudes where the solutions can be compared.    

\subsection{The Modified Morrison $\widetilde{G}$ Transform}
Application of the standard Morrison $G$ transform to plasma seismology faces the problem that for the Vlasov-Poisson problem the transform is k-dependent and is ideally adapted for the initial value problem and not for a boundary-value problem.  This problem can be circumvented, and the seismology problem solved, in the case of ion-acoustic waves at low frequency because the dielectric can be approximately expressed without the need of a Fourier transformation.
\citep{Skiff:2002a}
\begin{equation}
Imag(\epsilon)=-\pi C_s^2\frac{\partial f_0}{\partial v}
\end{equation}
Here we would like to propose a modification to the $G$-transform to make it applicable in situations where one would rather Fourier-transform on time than on space.  Consider the Vlasov-Poisson system again, but now we will replace $k$ with $\omega/v$ in our expression for the imaginary part of $\epsilon$.  We will also allow for the possibility of a velocity-independent normalization factor $N$.  
\begin{equation}
Imag(\tilde{\epsilon})=-\pi\frac{\omega_{p}^2 v^2}{\omega^2}\frac{\partial f_0}{\partial v}N
\end{equation}
The extra factors of $v$ seem to create a modification to the dispersion relation because the real part of $\tilde{\epsilon}$ is obtained by a Hilbert transform on $v$ of the imaginary part.
\begin{equation}
Real(\tilde{\epsilon})=1+Hilbert[Imag(\tilde{\epsilon})]_u
\end{equation}
However, repeated application of the well-known identity
\begin{equation}
Hilbert[vf]_u = \frac{1}{\pi}\int f dv + u*Hilbert[f]_u
\end{equation}
shows that the resulting dispersion relation is correctly that of the electron plasma wave (with an overall multiplicative normalization factor) if the normalization factor $N$ is chosen such that
\begin{equation}
N=(1-\frac{\omega_p^2}{\omega^2})^{-1}.
\end{equation}
This means that
\begin{equation}
\tilde{\epsilon}=\epsilon N
\end{equation}
and that a modified $\widetilde{G}$ transform based on $\tilde{\epsilon}$ will have the same form as the standard transform because $\tilde{\epsilon}$ is also analytic in the upper $v$-plane.  Here $\omega$ is taken as a real parameter, and in the end one substitutes $v=\omega/k$ with complex $k$ to obtain the dispersion relation. Starting from the linearized Vlasov equation with the Fourier transform on time
\begin{equation}
 v\frac{\partial f}{\partial x}-i\omega f+\frac{q}{m}E\frac{\partial f_0}{\partial v}=0
\end{equation}
and applying the inverse modified Morrison $\widetilde{G}$ transform one obtains
\begin{equation}
 u\frac{\partial g}{\partial x}-i\omega g=\frac{Imag(\tilde{\epsilon})\varepsilon_0}{\pi|\tilde{\epsilon}|^2q}(\frac{q}{\varepsilon_0}\frac{\partial n}{\partial z}+\frac{\omega^2}{u^2}E)
\end{equation}
with the perturbed electron density given by 
\begin{equation}
    n=\int f dv.
\end{equation}
Using the first spatial derivative of the Poisson equation to relate the density to the electric field we obtain
\begin{equation}
 u\frac{\partial g}{\partial x}-i\omega g=\frac{Imag(\tilde{\epsilon})\varepsilon_0}{\pi|\tilde{\epsilon}|^2q}(\frac{\partial_2 E}{\partial z^2}+\frac{\omega^2}{u^2}E).
\end{equation}
Both sides of this equation are set to zero if we assume that both $g$ and $E$ behave as $e^{ikx}$ with k determined by the CVK dispersion relation
\begin{equation}
    k=\omega/u.
\end{equation}
With this approach we can now solve the seismology problem which requires determination of the CVK spectra from data at a single spatial point.  For each frequency component we have the expression for the corresponding CVK spectrum
\begin{equation}
    g(u,x)_\omega=g(u,x=0)_\omega e^{i\omega x/u}
\end{equation}
with
\begin{equation}
    g(u,x=0)_\omega = \widetilde{G}_\omega [f(v,x=0)_\omega]
\end{equation}
where we have used $\widetilde{G}_\omega[f]$ to refer to the inverse modified Morrison $\widetilde{G}$ transform of $f$ at frequency $\omega$.
Using the fact that 
\begin{equation}
\int f dv = \int g du
\end{equation}
one can determine $n(x,t)$ directly from an expression for $g(x,u,t)$ by integration over $u$.  Alternatively one can obtain $f(x,v,t)$ using the modified Morrison $\widetilde{G}$ transform at each frequency followed by an inverse Fourier transform.  We will illustrate the implementation of plasma seismology starting from data produced  by a numerical simulation.

\section{Numerical Simulation}
\label{sec:sims}
We test the development of plasma seismology here using kinetic simulation data of Langmuir waves using the Nonlinear Vlasov-Poisson Code \T{VP} \citep{Howes:2017a}.  This code solves the 1D-1V Vlasov-Poisson equations for both species $s$ distribution functions $f_s(x,v,t)$ and the electrostatic potential $\phi(x,t)$,
\begin{equation}
\vspace*{-0.1in}
\frac{\partial f_s}{\partial t} + v \frac{\partial f_s}{\partial x} -
\frac{q_s}{m_s} \frac{\partial \phi}{\partial x} \frac{\partial
  f_s}{\partial v} =0,
\end{equation}
\begin{equation}
  \frac{\partial^2 \phi}{\partial x^2} = \frac{1}{\epsilon_0}\sum_s
\int_{-\infty}^{+\infty} dv \ q_s f_s,
\label{eq:vlasov}
\end{equation}
over an interval $[-L/2,L/2]$ using $n_x$ points in physical space $x$ and over $[-v_{max},v_{max}]$  using $n_v+1$ points in velocity space $v$. Spatial derivatives are computed using second-order, centered finite differencing with  periodic boundary conditions. Velocity derivatives are computed using second-order centered finite differencing, except for first-order differencing at $v=\pm v_{max}$.  The distribution function is advanced using a third-order Adams-Bashforth scheme. The maximum timestep for this explicit algorithm is constrained by the Courant-Friedrichs-Lewy stability criterion to be $(\Delta t)_{max} = \Delta x/\max(v_{max},\omega/k)$, where the maximum velocity is either the maximum resolved particle velocity $v_{max}$ or the maximum phase velocity of the Langmuir wave mode, $\omega/k$. The timestep used in the simulation is set to be some fraction $f_{CFL}$ of this maximum timestep, $\Delta t = f_{CFL} (\Delta t)_{max}$.  At each timestep, the Poisson equation is solved for the potential $\phi(x)$ using the Green's function solution.   The code may be run linearly or nonlinearly, and with or without dynamic ions.

We test the plasma seismology technique introduced here using a 1D-1V simulation of three Langmuir wave modes with $k \lambda_{de}=1/3, 1/2, 2/3$. The modes are initialized using an initial electron density variation $n_e(x,t_0)$ with random phases in each of the initialized wavenumber modes and are evolved forward in time \textit{nonlinearly}.  The full electron distribution function data cube $f(x,v,t)$ output from the simulation is used to compute both the standard Morrison $G$ transform using initial values at all positions at $t=t_0$, $f(x,v,t_0)$, and our modified version using a time series over $0 \le \omega_{pe} t \le 95.5$ at a single position $x_0$, $f(x_0,v,t)$.

\begin{figure}[t]
\resizebox{3.3in}{!}{\includegraphics{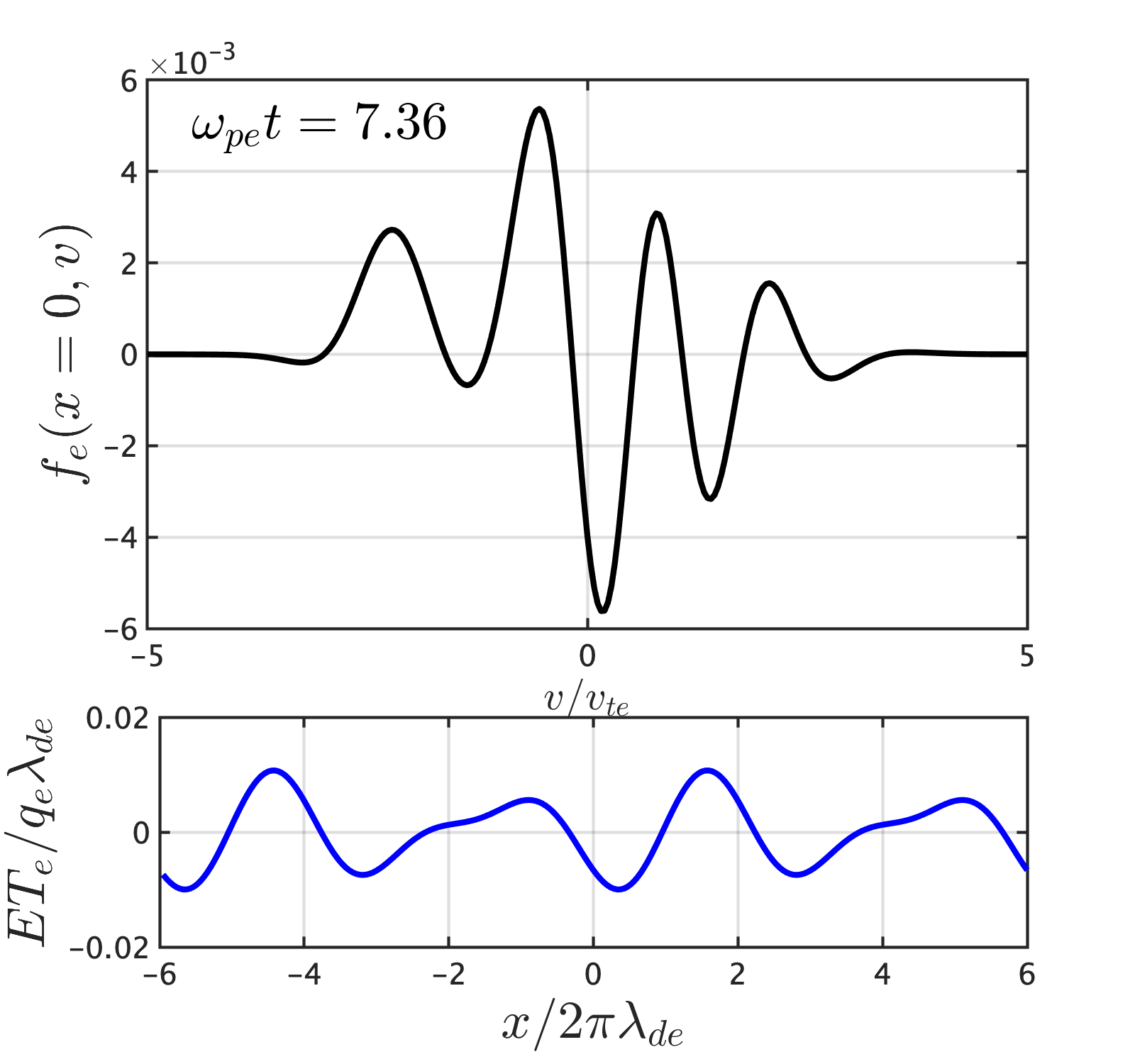}}
\resizebox{3.3in}{!}{\includegraphics{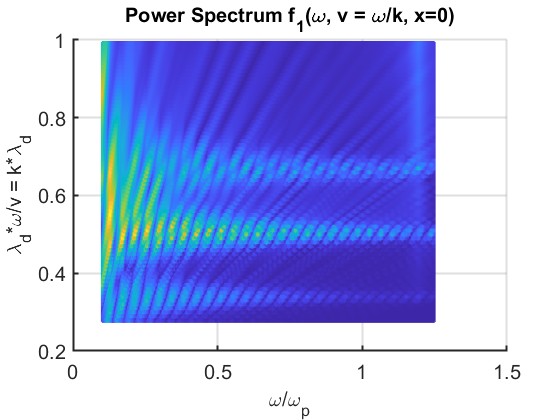}}
\caption{Sample 1D-1V \T{VP} simulation data, showing (a) the eVDF $f_e(x=0,\omega_{pe}t=7.36)$,
  (b) electric field $E(x)$, and (c) contours of $f_e$ on  the $(\omega,k=\omega/v)$ plane, clearly showing power in three modes (horizontal lines) at $k \lambda_{de}=1/3, 1/2, 2/3$. \label{fig:sims}}
\end{figure}

A sample of the data from this simulation is presented in \figref{fig:sims}, where we show (a) the eVDF $f_e(x_0,v,t_0)$ at position $x_0=0$ and time $\omega_{pe}t_0=7.36)$ and (b) the electric field over all positions $E(x,t_0)$.  It is clear from panels (a) and (b) that the variation in velocity and space is nontrivial, providing a discriminating test for the plasma seismology technique. We also plot (c) the frequency spectrum of the Langmuir modes in the simulation at $x_0$ by taking a Fourier transform in time of the eVDF at each velocity $v$, and plotting the results as a contour plot of power in normalized frequency $\omega/\omega_{pe}$ and normalized wavenumber $k \lambda_{de}$.  We obtain the wavenumber at each $(\omega,v)$ point through the relation $k=\omega/v$.  Note that the power in the three modes at $k \lambda_{de}=1/3, 1/2, 2/3$ is clearly evident in this plot; broadening of the power at each of these three wavenumbers arises because all of these modes are strongly damped, leading to broadening of the Fourier transformed signal in time.
\section{Results}
\label{sec:results}
Here we present the application of the standard Morrison $G$ transform  and the modified Morrison $\widetilde{G}$ transform to the 
nonlinear kinetic 1D-1V simulation of three Langmuir wave modes with $k \lambda_{de}=1/3, 1/2, 2/3$ using the Nonlinear Vlasov-Poisson code \T{VP}. 

From the simulation, we output at time series of the perturbed electron velocity distribution $f_1(x_0,v,t)$ at the single point $x_0=0$.  With this information as our input, the plasma seismology analysis is essentially the solution of a boundary value problem. First, we Fourier transform in time $t$ the distribution function $f_1(x_0,v,t)$ to obtain $\hat{f}_1(x_0,v,\omega)$.
Next, we perform the inverse modified Morrison $\widetilde{G}$ transform on this frequency spectrum,
\begin{equation}
    \hat{g}_1(x_0,u,\omega)= \widetilde{G}^{-1}\left[\hat{f}_1(x_0,v,\omega)\right](u).
\end{equation}
Note that $\omega$ may be considered a constant parameter in this transform, so the transform is computed for each $\omega$ value in $\hat{f}_1(x_0,v,\omega)$.

The complex coefficients of the inverse modified $\widetilde{G}$ transform given by $\hat{g}_1(x_0,u,\omega)$ are equivalent to the determining the Case-Van Kampen (CVK) spectrum of modes in the collisionless plasma \citep{Case:1959,VanKampen:1955}. 
Note that the solution is a function of $x$ in the inverse $G$ transformed space for each value of the frequency $\omega$  and velocity $u=\omega/k$, given by $\hat{g}_1(x,u,\omega)=\hat{g}_1(x_0,u,\omega) e^{-i(\omega/u) x} $.  

\begin{figure*}[t]
\begin{flushleft} 
\resizebox{3.5in}{!}{\includegraphics{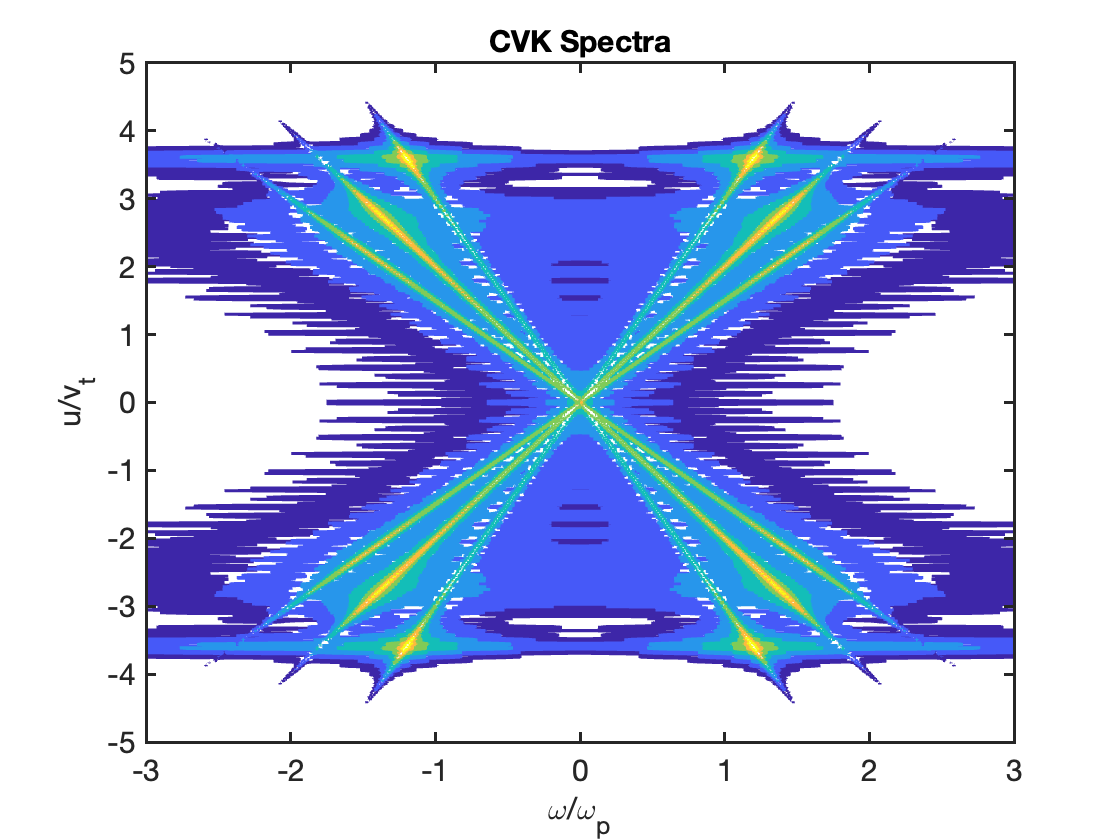}}
\resizebox{3.5in}{!}{\includegraphics{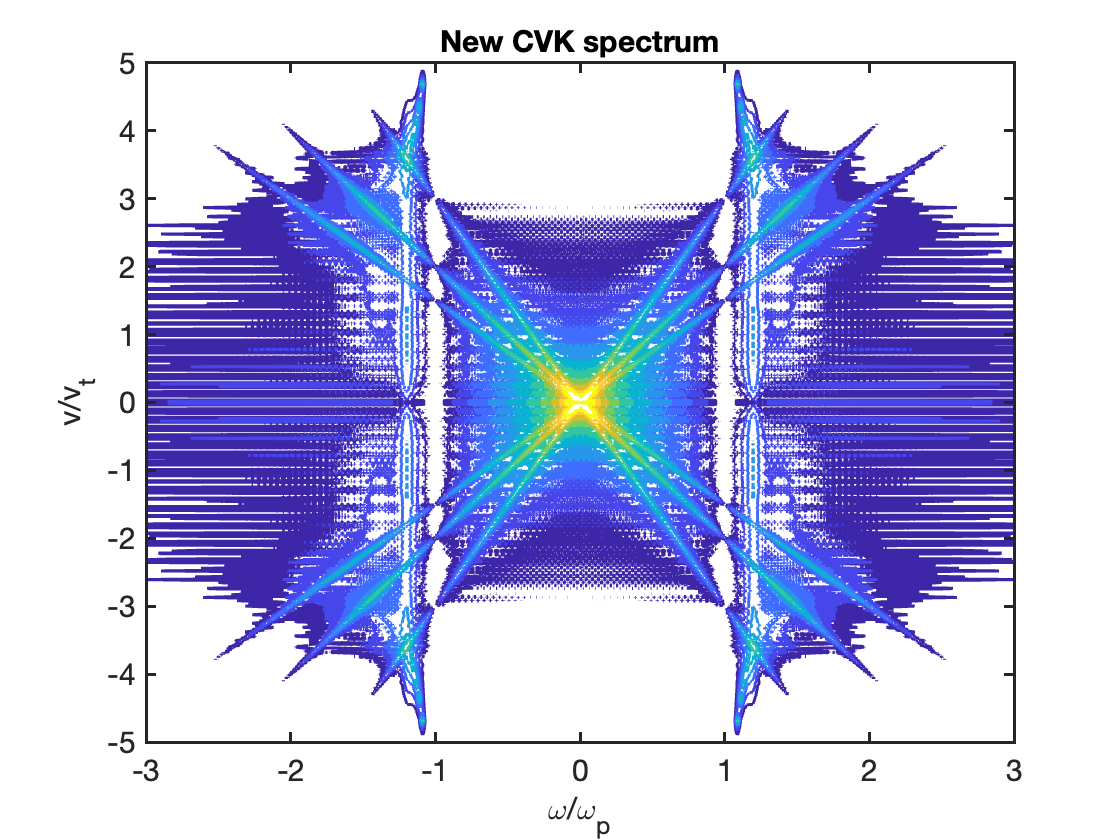}}\\ 
\resizebox{3.5in}{!}{\includegraphics{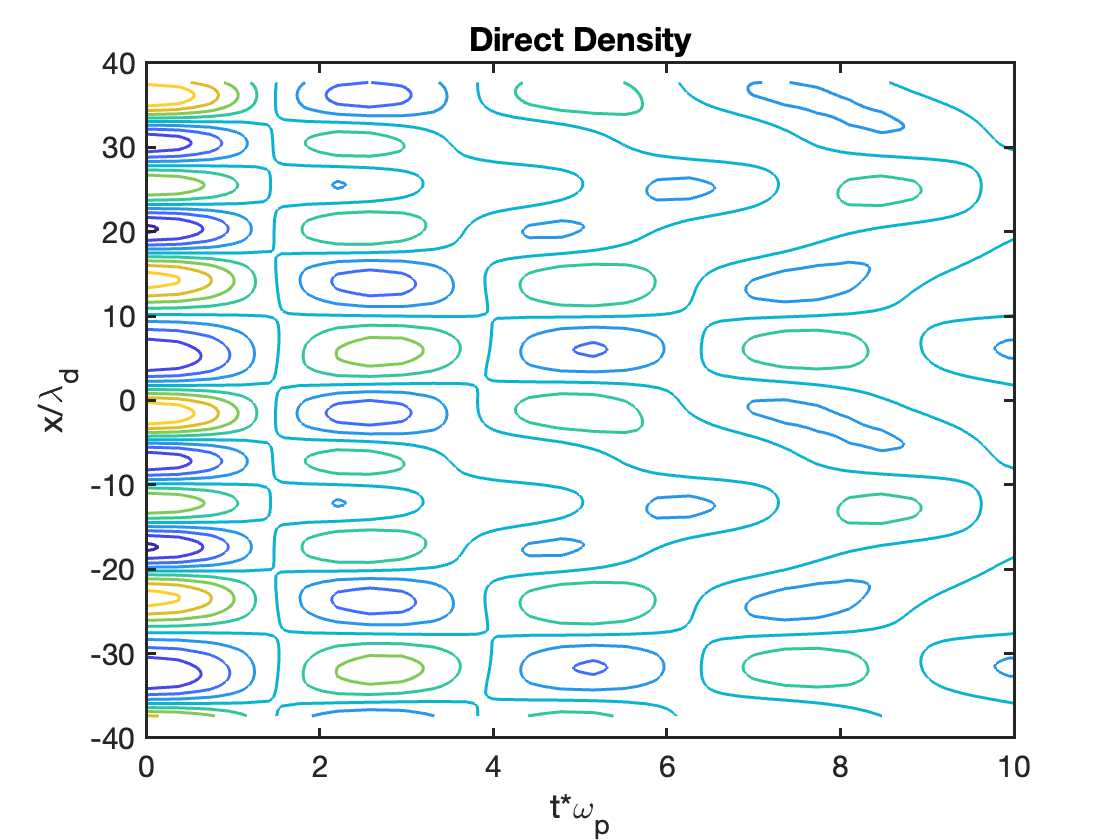}}
\resizebox{3.5in}{!}{\includegraphics{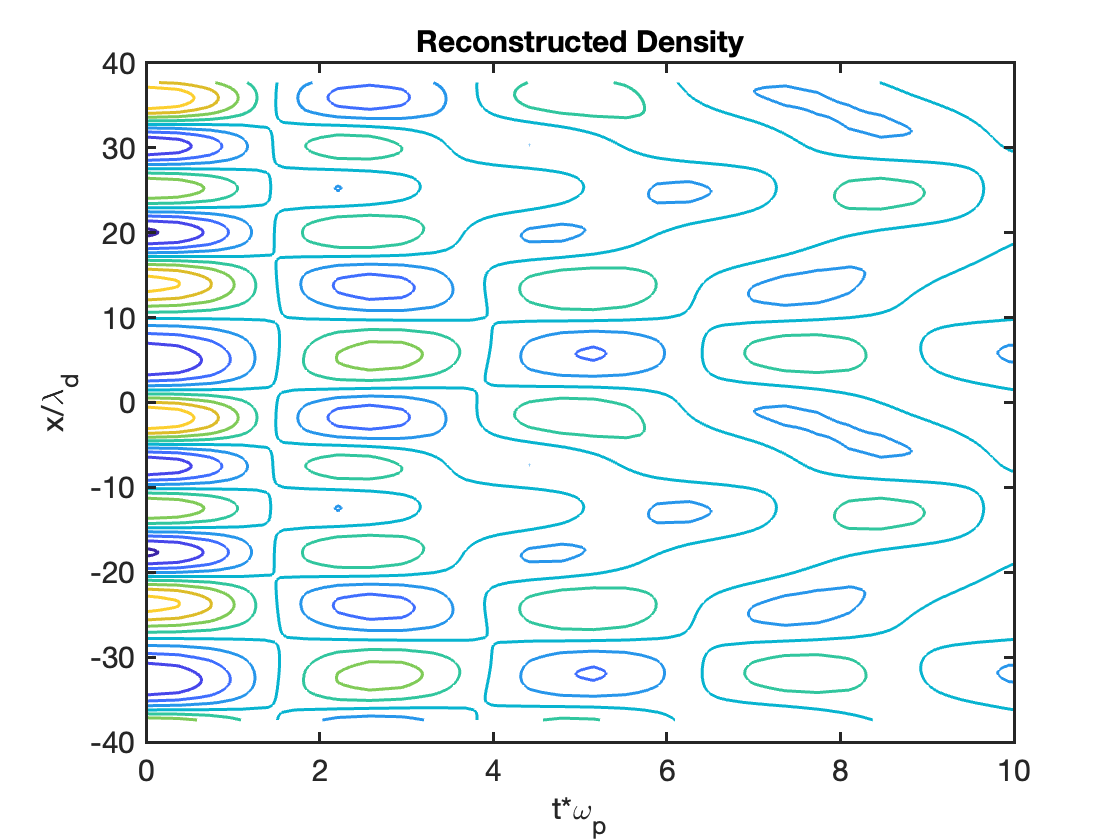}}\\ 
\vskip -5.2in
\vbox{ \hsize 6.5 in (a) \hspace{3.4 in} (b)}
\vskip 2.5in
\vbox{ \hsize 6.5 in (c) \hspace{3.4 in} (d)}
\vskip 2.3in
 \caption{ Preliminary results of boundary value formulation of the Morrison $G$ transform: Case-Van Kampen mode spectra from (a) the standard  Morrison $G$ transform computed using $f(x,v,t_0)$ and (b) our boundary value implementation of the Morrison $G$ transform computed using $f(x_0,v,t)$; the electron density variations $n_e(t,x)$ over all space and time (c) directly from the \T{VP} simulation and (d) computed using our modified Morrison $G$ transform.
 \label{fig:results}
}
\end{flushleft}
\end{figure*}

In \figref{fig:results}, we present the CVK spectrum from (a) the standard  Morrison $G$ transform (using the initial values $f_1(x,v,t=0)$) and (b) the modified Morrison $\widetilde{G}$ transform.  Note that each of the three wavenumbers initialized (with  $k \lambda_{de}=1/3, 1/2, 2/3$) generates a pair of the diagonal lines, corresponding to forward and backward propagating Langmuir waves with $u=\pm \omega/k$.  Note that the peak at $(\omega,u) \simeq (0,0)$ in Figure 4(b) arises from causality limitations due to the finite duration $\tau$ of our time series, but does not cause problems in the field reconstruction, as shown below. 

The next step is to perform the forward  modified Morrison $\widetilde{G}$ transform for each coefficient of this $u$ spectrum of modes back to $v$ space
    \begin{equation}
  \hat{f}(x_0,v,\omega)=\widetilde{G}\left[\hat{g}_1(x_0,u,\omega) e^{-i(\omega/u) x}\right](v)
\end{equation}
The final step is to inverse Fourier transform from frequency space $\omega$ back to time $t$ to obtain the final solution for the perturbed velocity distribution function \emph{at all positions and times}, $f_1(x,v,t)$.

From this perturbed eVDF, we can compute the density $n_e(x,t)$ and electric field $E(x,t)$ over all positions and times from the moments of $f_1(x,v,t)$.

In \figref{fig:results}, we compare the nontrivial temporal evolution of the electron density over the full spatial domain in $x$ from (c) direct output from the  \T{VP} simulation and (d) the reconstruction derived from our boundary value implementation using only the time series of the eVDF measurements $f_1(x_0,v,t)$ at $x_0=0$.  The reconstruction is extremely accurate, validating our boundary value formulation of the Morrison $G$ transform. This demonstration also shows that any variations in the Case-Van Kampen mode spectrum $\hat{g}_1(x_0,u,\omega)$ between the standard $G$ and modified  $\widetilde{G}$ transforms in \figref{fig:results}(a) and (b) does not cause any inaccuracies in the determination of the electron density (and thus the electric field) variation over space and time.

Note that a key step in our alternative boundary value formulation is to replace $k=\omega/u$ in the dielectric function, leading to a well-posed formulation of the modified Morrison $\widetilde{G}$ transform in terms of the (constant) parameter $\omega$ and the transform variable $u$. The excellent agreement between Figure 4 (c) and (d) for the electron density variations over all space and time, $n_e(t,x)$, validates our modified Morrison $\widetilde{G}$ transform for the case of electrostatic variations in the plasma.

\section{Limitations and Uncertainties}
\label{sec:limitations}
It is important to determine  the limitations and uncertainties associated with the modified Morrison $\widetilde{G}$ transform 
for reconstructing the spatial variation of the electric field from single-point measurements.
From theoretical considerations, we have identified that the uncertainty of the electrostatic electric field reconstruction using plasma seismology will depend on the following characteristics of the eVDF measurements  $f(x,v,t)$:
\begin{enumerate}
\item[] \hspace*{-0.4in} {\bf (i) eVDF Time Resolution and Time Series Duration}: The time resolution $\Delta t$ limits the maximum resolvable electric field frequency through the usual Nyquist frequency $\omega_{Ny}=\pi/\Delta t$, and the duration $\tau$ limits the resolution in frequency space $\Delta \omega= 2 \pi/\tau$. In addition, the maximum spatial extent of the plasma seismology determination of the electrostatic electric field will be limited by causality to locations where particles can reach over the measurement interval $\tau$, yielding $x_{max} \lesssim v_{max} \tau$.
\item[] \hspace*{-0.4in} {\bf (ii) Velocity Resolution}:  The velocity resolution $\Delta v$ limits the resolution in physical space $\Delta x$ through limitations on the wavenumber resolution through the relation $k = \omega/v$.
\item[] \hspace*{-0.4in} {\bf (iii) Effective Particle Velocity Limits}: The minimum and maximum measurable particle velocities $v_{min}$ and $v_{max}$ constrain the range of sensitivity of the technique to electrostatic fluctuations with characteristic linear phase velocities $v_{min} \lesssim \omega/k \lesssim v_{max}$. For application to spacecraft measurements, spacecraft charging effects effectively limit $v_{min}\simeq 0.5 v_{te}$ and the particle counting statistics limit  $v_{min}\lesssim 3 v_{te}$.
\item[] \hspace*{-0.4in} {\bf (iv) Collisionality}: Finite collisionality $\nu$ will limit the applicability of the Morrison $G$ transform, which is formally only applicable in the collisionless limit; practically, a finite collisionality $\nu \tau \gtrsim 1$ will erase the phase-space information carried by the particles that is necessary to reconstruct the electric field.
\end{enumerate}
Here we vary all of these parameters using \T{VP} simulations to determine how the uncertainties in the plasma seismology determinations of $E(x,t)$ depend on $\Delta t$, $\tau$, $\Delta v$,  $v_{min}$, $v_{max}$, and $\nu \tau$.

\subsection{Nonlinearity}
Since the standard Morrison $G$ transform and  modified Morrison $\widetilde{G}$ transform are both based on the \emph{linearized} Vlasov equation, one may wonder about the limitations of this approach to typical conditions in laboratory and space plasmas where nonlinear effects often arise. Several arguments suggest that, although the technique is based on linear theory, the deviations arising from any nonlinearity are likely to be small over the measurement duration $\tau$ used for the electric field reconstruction.  First, although the integro-differential Vlasov-Poisson equations include nonlinearity through the electric field term in the Vlasov equation, the Vlasov equation itself is linear in $f$. Therefore, one may define the mean velocity distribution $\langle f \rangle_\tau$ over a medium timescale,  here chosen to be the sampling duration $\tau$ to be used for the field reconstruction (typically about 10 linear wave periods, $\tau \simeq 10 T$).  Thus, the nonlinear Vlasov equation can be linearized about this equilbrium velocity distribution (constant over sampling duration $\tau$), yielding a linear equation for the evolution of the fluctuations $f_1=f-\langle f \rangle_\tau$. Although the well-known quasilinear evolution of the distribution function can lead to large amplitude changes of $f$ on long time scales $\mathcal{T} \gg \tau$ (much more than 10 wave periods), the amplitude of the perturbed velocity distribution is typically small relative to the medium-timescale equilibrium, $|f_1| \ll | \langle f \rangle_\tau|$.  Note that the medium-timescale equilibrium  $\langle f \rangle_\tau$ can deviate significantly from a Maxwellian, but the perturbations over the medium timescale will remain small.  Only if the field amplitudes reach large enough values to generate significant trapped particle populations, which can occur in particular situations (such as large-amplitude waves found in the quasiparallel foreshock region ahead of Earth's bow shock), are nonlinearities expected to foil the plasma seismology reconstruction.

Furthermore, in a turbulent plasma, the parallel electric fields that mediate strong wave-particle interactions occur at the ion kinetic scales and smaller; since turbulent amplitudes monotonically decrease with scale, the fluctuation amplitudes of the turbulence at the kinetic scales ($k_\perp\rho_i \sim 1$) will be small even in strong turbulence, yielding $|\delta \V{B} (k_\perp\rho_i \sim 1)|/B_0 \ll 1$.  Thus, a linear description the plasma response is likely to be accurate to lowest order on the medium timescales of the observation $\tau$.

The amplitude of the Langmuir waves, conveniently characterized by the spatial average of the magnitude of the perturbed density to the mean density, $\langle |\delta n|/n_0 \rangle_x$, controls the development of nonlinearity.  We can test this carefully by exploiting the \T{VP} code's capability to turn on or off the nonlinear evolution while varying  $\langle |\delta n|/n_0 \rangle_x$. 
\subsection{Wave Dispersion}
Langmuir waves are strongly dispersive, with phase velocities much larger than the electron thermal velocity, $\omega/k \gg v_{te}$ for $k \lambda_{de}\ll 1$; therefore, plasma seismology is expected to be limited to recovering  electric fields associated with Langmuir waves with $k \lambda_{de} \rightarrow 1$.  However, the electric fields with $k \lambda_{de} \rightarrow 1$ are the most important to study for the plasma dynamics and energy transport---such as collisionless damping of wave-like fluctuations---because they strongly interact with the particle velocity distributions, so we are likely to be able to capture the electric fields within this wavenumber range.  The ion acoustic waves analyzed experimentally using the an early version of the modified Morrison $\widetilde{G}$ transform \citep{Skiff:2002a} are non-dispersive, making that application much more simple; the dispersive Langmuir waves to be analyzed here provide a more stringent test of the plasma seismology technique. The \Alfven waves 
that dominated the dynamics in typical space and astrophysical plasmas are also non-dispersive in the long-wavelength limit and weakly dispersive at the ion kinetic length scales, so we anticipate that this extension of the modified Morrison $\widetilde{G}$ transform will be well behaved, similar to the ion acoustic waves in the laboratory \citep{Skiff:2002a}.

\subsection{Changes in Magnetic Field Direction and Large-Scale Gradients}
Since plasma seismology in a magnetized plasma samples particles that travel helically along the local magnetic field direction, an important question is whether the technique works if the magnetic field direction changes substantially over the measurement duration $\tau$.  Since the plasma seismology technique effectively inverts the perturbations to the particle velocity distribution arising from interactions with the parallel component of the electric field, we can obtain guidance from a particularly useful form of the energy equation in nonlinear gyrokinetics \citep{Huang:2024}, given by 
\begin{eqnarray}
  \label{eq:fpcingyro}
  \lefteqn{\frac{\partial w_s}{\partial t}
  + v_\parallel  \frac{\partial w_s }{\partial z}
+ \frac{T_{0s} c}{B_0 F_{0s}} \left[\langle \chi \rangle_{\V{R}_s},\frac{g_s^2}{2}\right]
    = } \\
   & &  v_\parallel \bhat \cdot  \left\langle q_s \V{E} \right\rangle_{\V{R}_s} g_s
   -   v_\parallel  \mu_s \bhat \cdot \nabla _{\V{R}_s}
   \left\langle \delta B_\parallel \right\rangle_{\V{R}_s} g_s \nonumber
\end{eqnarray}
where the phase-space energy density is $w_s = T_{s} g_s^2/(2F_{0s})$ and the complementary perturbed distribution function is given by  $g_s   = h_s  - \frac{q_s F_{0s}}{T_{0s}} \left\langle  \phi -  \frac{\V{v}_\perp \cdot \V{A}_\perp}{c}\right\rangle_{\V{R}_s}$, and $h_s$ is the perturbed gyrokinetic distribution function \citep{Howes:2006,Schekochihin:2009,Huang:2024}.  The right-hand side includes two nonlinear terms that describe collisionless wave-particle interactions: (i) the first term describes Landau damping by the component of the electric field along the local magnetic field direction, $E_\parallel=\V{E} \cdot \bhat$; (ii) the second term describes transit-time damping by the magnetic mirror force along the total magnetic field direction $\bhat = (\V{B}_0 + \delta \V{B}_1)/|\V{B}|$.  The nonlinearity arises in correcting from the  equilibrium  magnetic field to the total magnetic field direction using $\delta \V{B}_1$, but the parallel dynamics of the wave-particle interaction remains essentially unchanged.  In the context of plasma seismology, the intuition we gain from inspection of this equation is that the nonlinear corrections simply account for the locally perturbed direction of the total magnetic field.  Thus, since the individual particle motions follow helical trajectories about the local magnetic field direction, the particles sampled by plasma seismology simply map out the component of the electric field \emph{along} the total magnetic field direction.  Thus, the presence of changes in the local magnetic field direction do not significantly influence the physics of the wave-particle interactions, but the spatial profiles of the parallel electric field recovered from plasma seismology must be interpreted as following along the total magnetic field direction.

\section{Conclusion}
\label{sec:conc}
Here we introduce the innovative  \emph{Plasma Seismology} technique which uses measurements of the velocity distribution function (VDF) of particles at a single point in space to reconstruct the variation of the parallel electric field $E_\parallel$ over an extended spatial region.  The technique has the promise in the laboratory to use velocity distribution function (VDF) measurements to determine $E_\parallel$, even in cases where the direct measurement of $E_\parallel$ is not possible due to unattainable alignments of electric field double probes.  In a magnetized plasma, $E_\parallel$ plays a key role in collisionless wave-particle interactions via the Landau resonance, making it an important quantity to measure to understand plasma heating and particle acceleration. In the long term, a better determination of $E_\parallel$, using VDF measurements that can be made with existing spacecraft  instrumentation for  particle measurements (\emph{e.g.}, electrostatic analyzers), will improve our understanding of the wave-particle interactions in space plasmas. Such a new capability would provide valuable information needed to understand the dynamics of weakly collisional space plasmas, in particular phenomena with significant societal impacts, such as severe space weather events that can impact communication and navigation satellites as well as potentially cause severe damage to the electrical power grid \citep{NAS:2008}.  Ultimately, plasma seismology will enable the full exploitation of the information contained in the VDFs, information that is routinely measured by spacecraft but is often underutilized. 

We have reviewed the mathematical formulation of the standard Morrison $G$ transform, which solves the initial value problem of the linear kinetic evolution of the plasma.  We have introduced the  modified Morrison $\widetilde{G}$ transform, which provides a complementary approach that solves a boundary value problem using a time series of velocity distribution measurements measured at a single point in space.  This latter implementation has potentially great value because it can be applied in the laboratory and space environments using existing technology that measure the particle velocity distributions.

We have tested the  modified Morrison $\widetilde{G}$ transform using 1D-1V nonlinear kinetic simulations of multiple Langmuir waves, finding that the technique does an excellent job of reconstructing the spatial profile of the electron density perturbations in comparison to the direct results from the simulation.  We have characterized the limitations and uncertainties of the technique due to realistic limitations on the single-point measurement of the electron velocity distribution, including time resolution, velocity resolution, duration of the measurement, effective velocity limits, collisionality, and nonlinearity.


Although the original formulation of the Morrison $G$ transform focused on the electrostatic Vlasov-Poisson system \citep{Morrison:1992,Morrison:1994,Morrison:2000}, the primary application of plasma seismology will be to diagnose the parallel electric field in magnetized plasmas, such as heliospheric or laboratory plasmas. Heninger \& Morrison (2018) have formulated the specific mathematical form of the Morrison $G$ transform for the electrostatic gyrokinetic and drift-kinetic equations  \citep{Heninger:2018}.  Future extensions of this work aim to generate a boundary value formulation of plasma seismology in the electromagnetic limit and validate it using kinetic simulations of magnetized plasma turbulence.  

A complementary extension of this work, albeit one that requires a very different mathematical formulation, is to develop a similar technique that employs the Larmor motion of ions perpendicular to the magnetic field to map out the spatial variation of the electromagnetic fields in the plane perpendicular to the magnetic field.  A successful preliminary application of this idea in laboratory plasmas showed promising results \citep{Skiff:1992}, and further work to refine the technique and characterize the limitations and uncertainties would effectively add a perpendicular extension to the plasma seismology technique introduced here.

\begin{acknowledgments}
We acknowledge Phil Morrison for his analytical development of the $G$ Transform.
\end{acknowledgments}

\section*{Data Availability Statement}

The numerical simulation data presented here is publicly available at Zenodo.

\bibliography{abbrev2,gyro,solwind,gtransform23_refs,gtransform_space}

\end{document}